\documentclass[fleqn,12pt,twoside]{article}
\usepackage{espcrc1}

\usepackage{epsfig}
\usepackage[figuresright]{rotating}

\newcommand{\gtrsim}{\:\lower 0.4ex\hbox{$\stackrel{\scriptstyle >}
{\scriptstyle\sim}$}\:}
\newcommand{\lesssim}{\:\lower 0.4ex\hbox{$\stackrel{\scriptstyle <}
{\scriptstyle\sim}$}\:}

\title{The $r$-process: recent progress and needs for nuclear data}

\author{Y.-Z. Qian\address{School of Physics and Astronomy, 
        University of Minnesota, Minneapolis, MN 55455,\\
        United States of America}%
        \thanks{This work was supported in part by US DOE grants 
        DE-FG02-87ER40328 and DE-FG02-00ER41149.}
        }
       
\begin{document}

\maketitle

\begin{abstract}
Several nuclear physics issues essential to understanding the 
$r$-process are discussed. These include validity of the
waiting-point approximation, strength of closed neutron shells
in neutron-rich nuclei far from stability, and effects of neutrino 
interaction with such nuclei. The needs for nuclear data in 
resolving these issues are emphasized.
\end{abstract}

\section{INTRODUCTION}
The dominant mechanism for producing nuclei beyond the Fe group is
neutron capture. In particular, the rapid neutron capture process,
or the $r$-process, is responsible for approximately half the solar
abundances of nuclei with mass numbers $A>70$. The relative 
abundances of $r$-process nuclei in the solar system are shown in
Figure 1. This solar $r$-process abundance pattern ($r$-pattern)
has two prominent peaks at $A\sim 130$ and 195, respectively, and
provides an important basis for studying the $r$-process.

\begin{figure}[tb]
\begin{minipage}{1.5cm}
\makebox[1cm]{}
\end{minipage} 
\begin{minipage}{8cm}
\epsfig{file=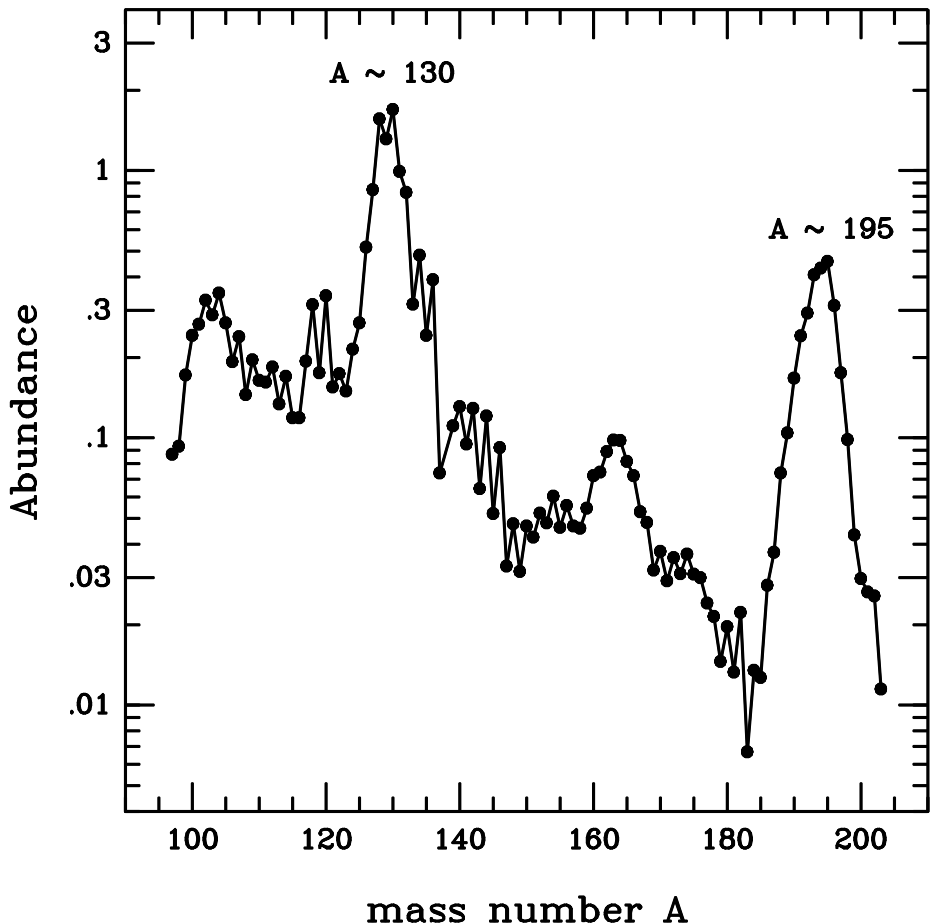,scale=0.7}
\end{minipage}
\caption{Solar $r$-process abundance pattern as derived in 
\cite{Kapp89}.}
\end{figure}

By definition, neutron capture occurs much more rapidly than
$\beta$ decay during the $r$-process. With many neutrons available
for capture by each seed nucleus, the nuclear flow quickly moves
toward the neutron-drip line. At some point, the separation energy
of the next neutron to be captured becomes so small that it will be
quickly disintegrated by the photons in the $r$-process environment.
At this so-called ``waiting point,'' the nuclear flow is characterized
by a tug of war between neutron capture and photo-disintegration.
Heavier nuclei cannot be produced until the waiting-point nucleus
$\beta$ decays to its daughter with a higher proton number.
Then the tug of war repeats at the next waiting point. Through such
a series of neutron capture and $\beta$ decay, very neutron-rich
nuclei far from stability are produced. Clearly, at a given proton
number $Z$, the abundance is concentrated in the corresponding
waiting-point nucleus. The more slowly this nucleus $\beta$ decays,
the more abundant it is. As nuclei with magic neutron numbers
$N=82$ and 126 have (relatively)
very slow $\beta$-decay rates, abundance peaks
are produced at these nuclei. After neutron capture ceases, these
nuclei successively $\beta$ decay to stability and give rise
to the peaks at $A\sim 130$ and 195 in the solar $r$-pattern
(see Fig. 1).

\section{WAITING-POINT APPROXIMATION AND NUCLEAR PROPERTIES}
The tug of war between neutron capture and photo-disintegration 
during the $r$-process results in a so-called ``$(n,\gamma)
\rightleftharpoons(\gamma,n)$ equilibrium,'' for which the abundances
$Y(Z,A+1)$ and $Y(Z,A)$ of two neighboring nuclei $(Z,A+1)$ and $(Z,A)$
in the same isotopic chain satisfy (see e.g., \cite{Qian03})
\begin{eqnarray}
{Y(Z,A+1)\over Y(Z,A)}&=&{n_n\langle v\sigma_{n,\gamma}(Z,A)\rangle
\over\lambda_{\gamma,n}(Z,A+1)}\nonumber\\
&=&n_n\left({2\pi\hbar^2\over m_ukT}\right)^{3/2}
\left({A+1\over A}\right)^{3/2}{G(Z,A+1)\over 2G(Z,A)}
\exp\left[{S_n(Z,A+1)\over kT}\right],
\label{nge}
\end{eqnarray}
where $n_n$ is the neutron number density of the $r$-process environment, 
$\langle v\sigma_{n,\gamma}(Z,A)\rangle$ denotes the product of neutron 
velocity and capture cross section averaged over a thermal distribution 
of temperature $T$, $\lambda_{\gamma,n}(Z,A+1)$ is the photo-disintegration 
rate at the same temperature, $\hbar$ is the Planck constant, 
$m_u$ is the atomic mass unit, $k$ is the Boltzmann constant, 
$G(Z,A)$ is the nuclear partition function, 
and $S_n(Z,A+1)$ is the neutron separation energy. If $S_n$
monotonically decreases with increasing $A$ over an isotopic chain, then
Equation (\ref{nge}) indicates that the most abundant nucleus in this 
chain is the nucleus with an $S_n$ approximately equal to
\begin{eqnarray}
\bar S_n&=&kT\ln\left[{2\over n_n}
\left({m_ukT\over 2\pi\hbar^2}\right)^{3/2}\right]\nonumber\\
&=&\left({T\over 10^9\ {\rm K}}\right)\left\{2.79+0.198
\left[\log\left({10^{20}\ {\rm cm}^{-3}\over n_n}\right)
+{3\over 2}\log\left({T\over 10^9\ {\rm K}}\right)\right]\right\}\ 
{\rm MeV}.
\label{sn}
\end{eqnarray}
However, pairing leads to odd-even staggering superposed on a general
trend of decreasing $S_n$ with increasing $A$ for an isotopic 
chain. In contrast, the behavior of the two-neutron separation energy 
$S_{2n}$ is essentially monotonic. Consequently, the most abundant nucleus 
in an isotopic chain has an even $N$ and a two-neutron separation energy
$S_{2n}\approx 2\bar S_n$ \cite{Gor92}. As $\bar S_n$ only depends on 
$n_n$ and $T$ of the $r$-process 
environment, it can be seen that the abundances in 
$(n,\gamma)\rightleftharpoons(\gamma,n)$ equilibrium are concentrated in 
a set of even-$N$ nuclei that have approximately the 
same two-neutron separation energy $S_{2n}\approx 2\bar S_n$. These are the
waiting-point nuclei.

If $(n,\gamma)\rightleftharpoons(\gamma,n)$ 
equilibrium is indeed obtained during the $r$-process, then there is no need 
to follow neutron capture and photo-disintegration reactions in an 
$r$-process calculation. This so-called ``waiting-point approximation'' is 
valid only when the rates of neutron capture and photo-disintegration are 
much faster than those of $\beta$ decay for the nuclei involved in the
$r$-process. The photo-disintegration rate for nucleus $(Z,A+1)$
is related to the neutron capture cross section for nucleus $(Z,A)$ by
detailed balance as described in Equation (\ref{nge}).
Based on a range of theoretical input for neutron capture cross sections, 
nuclear masses, and $\beta$-decay rates, it was estimated that the
waiting-point approximation
is valid for $n_n\gtrsim 10^{20}$ cm$^{-3}$ at $T=2\times 10^9$ K and for 
$n_n\gtrsim 10^{28}$ cm$^{-3}$ at $T=10^9$ K \cite{Gor96}. 

The derivation of the constraints on $n_n$ and $T$ can be illustrated by 
considering production of the peak at $A\sim 195$. The expected 
waiting-point nuclei giving rise to this peak should include $^{195}$Tm 
with $N=126$. For $^{195}$Tm to be a waiting-point nucleus at least requires
\begin{equation}
\lambda_{\gamma,n}({^{195}{\rm Tm}})>\lambda_\beta({^{195}{\rm Tm}}),
\ n_n\langle v\sigma_{n,\gamma}({^{195}{\rm Tm}})\rangle>\lambda_\beta
({^{195}{\rm Tm}}),
\end{equation}
where $\lambda_\beta({^{195}{\rm Tm}})$ is the $\beta$-decay rate of
$^{195}$Tm.
Calculations give $\langle v\sigma_{n,\gamma}({^{194}{\rm Tm}})\rangle
=4\times 10^{-19}$ cm$^3$ s$^{-1}$, 
$\langle v\sigma_{n,\gamma}({^{195}{\rm Tm}})\rangle=7\times 10^{-22}$ 
cm$^3$ s$^{-1}$ \cite{Cow91}, $S_n({^{195}{\rm Tm}})=4.2$ MeV, and
$\lambda_\beta({^{195}{\rm Tm}})=10$ s$^{-1}$ \cite{Moll97}. Based on these
theoretical results, the waiting-point approximation may be used for
production of the $A\sim 195$ peak when $T>1.4\times 10^9$ K and 
$n_n>1.4\times 10^{22}$ cm$^{-3}$.

Clearly, the validity of the waiting-point approximation depends on both
astrophysical conditions in the $r$-process environment and nuclear
properties. If valid, this approximation greatly simplifies $r$-process
calculations and reduces the needs for cross sections of neutron capture
on neutron-rich nuclei far from stability. In view of this nice feature,
it is crucial that the validity of the waiting-point
approximation be established by experimental data. For example, the 
constraints on $n_n$ and $T$ for applying the waiting-point approximation 
to production of the $A\sim 195$ peak discussed above should be rederived
based on measurements of the neutron capture cross sections for 
$^{194}$Tm and $^{195}$Tm and the neutron separation energy
and $\beta$-decay rate for $^{195}$Tm.

\section{$r$-PATTERN RESULTING FROM CLASSICAL CALCULATIONS AND STRENGTH OF 
CLOSED NEUTRON SHELLS}
Classical $r$-process calculations \cite{See65} assume
$(n,\gamma)\rightleftharpoons(\gamma,n)$ equilibrium, which selects a set
of waiting-point nuclei with $S_{2n}\approx 2\bar S_n$. The value of
$\bar S_n$ is determined by $n_n$ and $T$ of the $r$-process environment
[see Eq. (\ref{sn})]. The relative abundances of the waiting-point 
nuclei are controlled by their $\beta$-decay rates and the duration of the
$r$-process $t_r$. A specific abundance pattern is produced for a set of 
$n_n$, $T$, and $t_r$ based on the nuclear physics input of two-neutron 
separation energies and $\beta$-decay rates for the relevant nuclei. The
goal of classical $r$-process calculations is to reproduce the solar
$r$-pattern by a superposition of patterns produced with a range of
$n_n$, $T$, and $t_r$. This approach succeeds in reproducing the gross 
features of the solar $r$-pattern, especially the peaks at $A\sim 130$ 
and 195. However, most such calculations severely underproduce the nuclei 
below these peaks (see e.g., \cite{Kra93}). These deficiencies can be 
traced to the astrophysical conditions and nuclear physics input assumed 
in such calculations.

\begin{figure}[tb]
\begin{minipage}{1.8cm}
\makebox[1cm]{}
\end{minipage} 
\begin{minipage}{8cm}
\epsfig{file=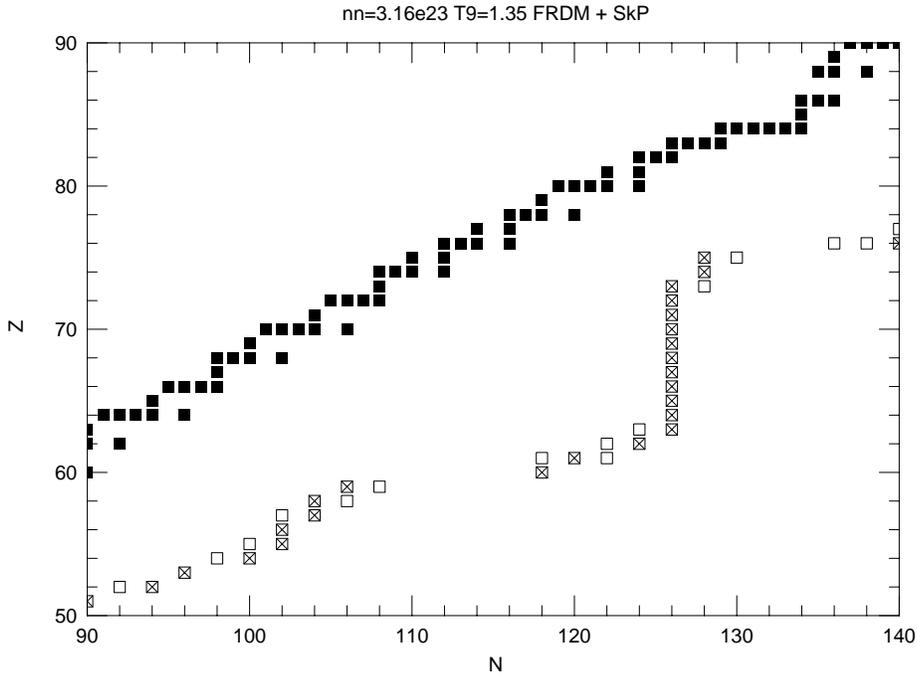,angle=90,scale=0.5}
\end{minipage}
\caption{Waiting-point nuclei (open and crossed squares) with 
$S_{2n}\approx 5.8$ MeV based on a typical mass formula. The stable 
nuclei are shown as filled squares for comparison. Courtesy Friedel
Thielemann.}
\end{figure}

To produce a component of the solar $r$-pattern, classical calculations
assume a fixed set of $n_n$ and $T$, which then choose a fixed set of
waiting-point nuclei with approximately the same $S_{2n}$. As the
waiting-point nuclei are far from stability, $S_{2n}$ has to be calculated 
from a mass formula for essentially all these nuclei. The waiting-point 
nuclei with $S_{2n}\approx 5.8$ MeV 
corresponding to $n_n=3.16\times 10^{23}$ 
cm$^{-3}$ and $T=1.35\times 10^9$ K are shown as open and crossed squares
in Figure 2 along with the stable nuclei shown as filled squares. It can be
seen that there is a large mass gap for the waiting-point nuclei below the
magic number $N=126$. This can be understood as follows.
In general, at a given $N$, $S_{2n}$ increases with increasing $Z$ while
at a given $Z$, it decreases with increasing $N$. Typical mass formulae
predict that $S_{2n}$ decreases very slowly as $N$ increases toward the
magic numbers. Consequently, when $Z$ increases from 59 to 60, $N$ has to
increase from 108 to 118 in order to reach the same 
$S_{2n}\approx 5.8$ MeV.
This results in a gap of 11 mass units for the waiting-point nuclei below 
the magic number $N=126$ and causes the severe underproduction below the
$A\sim 195$ peak. The deficiency below the $A\sim 130$ peak can be similarly
explained.

The severe underproduction of nuclei below the peaks at $A\sim 130$ and 195
in classical calculations suggest that the mass formulae used in these
calculations may be inadequate or that the astrophysical conditions are not
treated properly by these calculations. For example, these deficiencies can
be alleviated by quenching the strength of closed neutron shells in the
mass formulae \cite{Chen95}. On the other hand, $n_n$ and $T$ decrease with
time during a realistic $r$-process. So the waiting-point nuclei are not 
fixed as assumed in classical calculations. It was shown that the deficiency 
below the $A\sim 195$ peak can be removed by taking into account the time
evolution of $n_n$ and $T$ \cite{Fre99}. Clearly, to fully understand the
cause for the underproduction of nuclei below the peaks at $A\sim 130$ and 
195 in classical calculations, we need both accurate mass measurements for
nuclei far from stability and detailed description of the $r$-process
environment.

\section{$r$-PROCESS SITE AND EFFECTS OF NEUTRINO INTERACTION}
The $r$-process is commonly associated with core-collapse supernovae or
neutron star mergers. In the supernova model, neutrinos emitted from a
nascent neutron star drives a wind by heating the material above the
neutron star. The wind material is initially composed of neutrons and 
protons. As it expands away from the neutron star and cools, neutrons and
protons combine to produce seed nuclei. If many neutrons are left over for
each seed nucleus produced, then an $r$-process can occur in this 
neutrino-driven wind (see e.g., \cite{Woo92,Mey92,Tak94,Woo94}). However,
this model currently has sever difficulty in producing the nuclei with
$A>130$ (see e.g., \cite{Wit94,Qian96,Hof97,Thom01}). In the neutron star
merger model, the $r$-process is supposed to occur in the ejecta from an
old neutron star that is disrupted during the merger. By treating the
neutron-richness of the ejecta as a free parameter, it was shown that
an $r$-process can occur to produce the nuclei with $A>130$ \cite{FRT99}.
In fact, when such nuclei are produced in the neutron star merger model,
the heaviest nucleus fissions, thereby providing
fission fragments as new seed
nuclei to capture neutrons. This results in fission cycling with major
production of $A>130$ nuclei only \cite{See65,FRT99}.

The above summary of $r$-process models appears to suggest that supernovae
produce the nuclei with $A\lesssim 130$ while neutron star mergers produce
those with $A>130$. However, observations of abundances of Fe and 
$r$-process elements in old stars show that supernovae rather than neutron 
star mergers are the major source for $r$-process nuclei with $A>130$. 
Only supernovae can contribute Fe to stars. If supernovae also 
contribute $r$-process elements with $A>130$ such as Eu, then Eu should be
observed along with Fe in old stars. In contrast, if neutron star mergers
are the major source for Eu, then Eu should not be observed in very old 
stars that received Fe contributions from only a small number of supernovae.
This is because neutron star mergers are $\sim 10^3$ times less frequent 
than supernovae in the Galaxy. Many supernovae must have already occurred 
to provide Fe to old stars before neutron star mergers could provide Eu.
As $r$-process elements with $A>130$ such as Eu have been observed in stars
with Fe abundances corresponding to contributions from very few supernovae,
neutron star mergers can be ruled out as the major source for $r$-process
nuclei with $A>130$ \cite{Qian00,Arg03}.

The above discussion leads to a rather unsatisfactory situation: 
observations favor supernovae as the major source for $r$-process nuclei
with $A>130$ but current supernova models have difficulty in producing 
these nuclei. A crucial step in resolving this dilemma is to find some 
independent observational signatures that also support supernovae as the 
major source for these nuclei. For example, interaction of supernova 
neutrinos with the nuclei produced by the $r$-process may 
have left some signatures in the $r$-pattern. Supernova neutrinos have
average energies of 10--25 MeV. So neutrino interaction can highly excite 
the neutron-rich $r$-process nuclei, which can then deexcite through 
neutron emission. It was shown that the solar $r$-process abundances of
the nuclei with $A=183$--187 can be completely accounted for by 
neutrino-induced neutron emission from the nuclei in the $A\sim 195$ peak
(see Fig. 3) \cite{Qian97,Hax97}. In addition, supernova neutrinos may 
induce fission of the $r$-process nuclei in and above the $A\sim 195$ peak.
This would produce fission fragments below and above $A=130$ but very few
fragments with $A\sim 130$, which can account for the $r$-patterns observed
in two old stars \cite{Qian02}.

\begin{figure}[tb]
\begin{minipage}{1.5cm}
\makebox[1cm]{}
\end{minipage} 
\begin{minipage}{8cm}
\epsfig{file=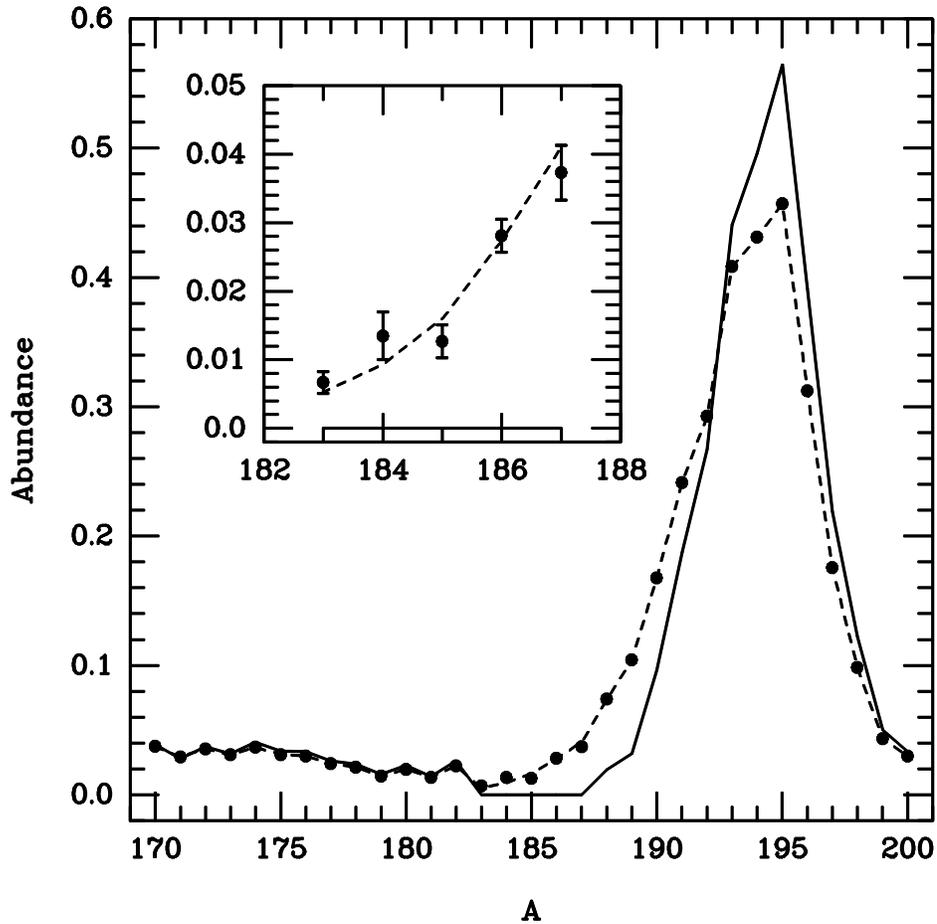,scale=0.7}
\end{minipage}
\caption{Production of nuclei by neutrino-induced neutron emission.
The abundances before and after neutrino processing are given by the solid 
and dashed curves, respectively. The filled circles (some with error bars) 
give the solar $r$-process abundances.}
\end{figure}

\section{CONCLUSIONS}
In conclusion, a full understanding of the $r$-process depends on a lot of
nuclear physics input. Neutron separation energies and capture cross 
sections and $\beta$-decay rates for some key nuclei such as $^{195}$Tm are 
needed to establish the validity of the waiting-point approximation. 
Accurate mass measurements for extremely neutron-rich nuclei are needed to 
quantify the effects of neutron shell strength on production
of the nuclei below the $A\sim 130$ and 195 peaks. Finally, branching
ratios of neutron emission and fission and fission yields for 
highly-excited neutron-rich nuclei are needed to understand the effects of 
supernova neutrinos on the $r$-pattern.

\end{document}